\documentclass[12pt]{iopart}

\newtheorem{df}{~~~{\bf Definition}}[section]
\newtheorem{thm}[df]{~~~{\bf Theorem}}  
  
\newtheorem{lm}[df]{~~~{\bf Lemma}}

\newcommand{\qed}{\raisebox{.6ex}{\fbox{\rule[0.0mm]{0mm}{0.8mm}}}}

\makeatletter
 \renewcommand{\theequation}{%
       \arabic{section}.\arabic{equation}}
 \@addtoreset{equation}{section}
\makeatother

\begin{document}

\title[The power decay of the survival probability at long times]
{The various  power decays of the survival probability \\
at long times for free quantum particle}

\author{Manabu Miyamoto
}

\address{Department of Physics, Waseda University, 3-4-1 Okubo, 
Shinjuku, Tokyo 169-8555,  Japan \\ 
E-mail: miyamo@hep.phys.waseda.ac.jp }
\date{\today}

\begin{abstract} 
The long time behaviour of the survival probability of initial state 
and its dependence on the initial states are considered, 
for the one dimensional free quantum particle. 
We derive the asymptotic expansion of the time evolution operator at long times, 
in terms of the integral operators. 
This enables us to obtain the asymptotic formula 
for the survival probability of the initial state 
$\psi (x)$, which is assumed to decrease sufficiently rapidly at large $|x|$. 
%
We then show that the behaviour of the survival probability at long times 
is determined by that of the initial state $\psi$ at zero momentum $k=0$. 
Indeed, it is proved that 
the survival probability can exhibit the various power-decays like $t^{-2m-1}$ 
for an arbitrary non-negative integers $m$ as $t \rightarrow \infty $, 
corresponding to the initial states 
with the condition $\hat{\psi} (k) = O(k^m )$ as $k\rightarrow 0$. 
%
\end{abstract}




\maketitle


\section{Introduction} \label{sec:1}

\setcounter{df}{0}

The decaying quantum systems 
such as an $\alpha$-decaying nucleus are often described 
by the survival probability of initial state, 
which is the probability to find the initial state in the state at a later time. 
One of the remarkable properties of the survival probability 
is its power decay law at long times. 
This is a mathematically predicted nature for the systems which possess 
the continuous energy spectrum bounded from below \cite{Fonda}. 
It is actually seen in many models 
(see, e.g.  \cite{Nakazato,Garcia,Muga}  
and the references therein). 
On the other hand, there still remains 
the difficulty in observing such power decays in the real experiments 
\cite{Greenland,Norman,Nicolaides}. 
Hence, the further theoretical and experimental investigations 
of the power decay law are much required.

One of the fundamental and important models that exhibits 
the power decay law is the free-particle system,  from which 
we can gain many insights into the dynamics 
of the survival probability and also the spatial wave packet. 
Recently, another aspect in the power-law regime at long times 
is revealed for the one dimensional free particle system, 
in connection with the initial states.  
As we known, the spatial wave packets for this system are expected to decay 
like $t^{-1/2}$ at long times $t$. 
We are assured of such a decay for the Gaussian wave packet. 
However, it is not necessarily valid for an arbitrary initial state. 
In fact, if we take a spatial power-law wave packet as an initial state, 
the ``anomalous decay'' of its maximum can occur with the form 
$t^{-\alpha/2}$ ($1/2 < \alpha < 1$) \cite{Unnikrishnan,Lillo,Mendes}. 
This is obviously slower than the well-known $t^{-1/2}$ decay. 
The faster decay than $t^{-1/2}$ is also studied 
for the initial wave packets which vanish at zero-momentum,  
in association with the dwell time \cite{Damborenea} 
%
and the time operator (see \cite{Miyamoto} and Appendix\ A). 
The latter refers to the time evolution of the survival probability. 
Therefore, the asymptotic decay form of the wave packet for the free particle system 
depends on the initial states in a considerable way. 
However, we seem still not to get a clear perspective 
of this new aspect of the power decay law. 
Our aim in the present paper is 
to find the strict condition of the initial states 
for the various power decays in the one dimensional free particle system, 
and to clarify the underlying mechanism for such decays. 
In particular, we restrict ourselves to the survival probability 
which will show the various power decays 
as same as the spatial wave packet.

To this end, we introduce a systematic approach which consists of the two procedures. 
We first derive in \sref{sec:6} the asymptotic expansion of the time evolution operator 
$\exp(-\rmi tH_0)$ as $t \rightarrow \infty$, where $H_0$ is the free Hamiltonian 
for the one dimensional free particle system. It is formally written as 
\begin{equation}
\exp(-\rmi tH_0) = \pi^{-1}  \sum_{j=0} ^{\infty} 
(-1)^{j-1} \Gamma (j+1/2) (\rmi t)^{-j-1/2} ~ G_{2j} , 
\label{eqn:1.100}
\end{equation}
where $\Gamma (z)$ is the gamma function, and $G_{2j}$'s are some integral operators. 
This kind of the asymptotic expansion was already developed by 
Rauch \cite{Rauch}, Jensen and Kato \cite{Jensen}, 
and Murata \cite{Murata} with the detailed analyses 
(see also a recent comment by Amrein \cite{Amrein}). 
Their results concern the time evolution operator 
for the systems with a short-range potential $V(x)$. 
The asymptotic expansion (\ref{eqn:1.100}) 
for the free particle system is not 
evaluated by the authors mentioned above, 
however, it can be achieved without difficulty, 
following their technique for the potential system. 
The survival probability of the initial state $\psi$ 
is defined by the square modulus of the survival amplitude of $\psi$, that is 
\[
\langle \psi , \exp(-\rmi tH_0) \psi \rangle . 
\]
We see from (\ref{eqn:1.100}) that 
$\langle \psi , \exp(-\rmi tH_0) \psi \rangle =O(t^{-1/2})$ 
only if $\langle \psi , G_0 \psi \rangle \neq 0$. 
In other words, if the following condition, 
\begin{equation}
\langle \psi , G_{2j} \psi \rangle = 0 ,~~~ j=0,1,\ldots, m-1, 
\label{eqn:1:105}
\end{equation}
holds for some integer $m$, we can obtain another asymptotic decay form of 
$\langle \psi , \exp(-\rmi tH_0) \psi \rangle$, $t^{-m-1/2}$, 
faster than $t^{-1/2}$. 
Thus, our remaining procedure is to interpret 
the condition (\ref{eqn:1:105}) 
as the behaviour of the initial state $\psi$ in, e.g., the position or momentum space. 
This is achieved in \sref{sec:7}, and 
we will finally obtain a remarkable conclusion: 
if the initial state $\psi$ behaves like $\hat{\psi} (k) = O(k^{m})$ 
at zero momentum $k=0$, then 
\begin{equation}
|\langle \psi, \exp(-\rmi tH_0) \psi \rangle |^2 = 
\pi^{-1} t^{-2m-1} \Gamma (m+1/2)^2 
|\langle \psi , G_{2m} \psi \rangle |^2 + o(t^{-2m-1} ) ,   
\label{eqn:1.110}
\end{equation}
as $t \rightarrow \infty$, 
where $\psi$ is assumed to decrease sufficiently rapidly at large $|x|$, 
but otherwise arbitrary. 
Hence, the behaviour of the initial $\psi$ at zero momentum, i.e., 
$\hat{\psi} (k) = O(k^{m})$, 
completely characterizes the asymptotic decay form of the survival probability of $\psi$. 
This fact is also expected 
from the results on the decays faster than $t^{-1/2}$ 
for the one dimensional free particle system \cite{Damborenea,Miyamoto}.

The organization of the paper is as follows. 
We first consider in \sref{sec:2} 
the asymptotic behaviour of 
the free resolvent $(H_0 -z)^{-1}$ 
at small and large energies.  
This study is necessary to the proof of 
Theorem\  \ref{thm:asymptotic evolution} in \sref{sec:6},  
where the asymptotic expansion (\ref{eqn:1.100}) is derived. 
The derivation essentially follows the method used by Jensen and Kato \cite{Jensen}. 
%
%
To derive the asymptotic formula (\ref{eqn:1.110}) for the survival probability, 
it is enough to derive that for the survival amplitude. 
The latter is accomplished in Theorem\  \ref{thm:dep on momentum} in \sref{sec:7}. 
Concluding remarks are given in \sref{sec:8}.



\section{The free resolvent 
in one dimension} 
\label{sec:2}

\setcounter{df}{0}

We here define the free Hamiltonian in one dimension 
$H_0 := P^2$, where 
$P$ is the momentum operator defined by $P:=-\rmi D_x$, 
$D_x$ being the differential operator on $L^2 ({\bf R})$. 
Then, the free resolvent $R_0 (z):=(H_0 -z)^{-1}$ is 
explicitly represented as 
an integral operator on $L^2 ({\bf R})$ \cite{RS2-1} 
\begin{equation}
(R_0 (z) \psi ) (x) = -\frac{1}{2\rmi z^{1/2}} 
\int_{{\bf R} } \exp(\rmi z^{1/2} |x-y|) \psi (y) \rmd y , 
\label{eqn:2.10}
\end{equation}
for all $\psi \in L^2 ({\bf R})$, where 
$z$ belongs to ${\bf C}_+ := \{ z \in {\bf C} ~|~ {\rm Im }\ z >0 \}$, 
and ${\rm Im }\ z^{1/2} > 0$. 
The resolvent $R_0 (z) $ is analytic in $z$. 
We, however, intend to regard it as an operator which maps 
$L^{2,s} ({\bf R})$ to $L^{2,-s^{\prime}} ({\bf R})$ 
for positive $s$ and $s^{\prime}$. 
Here $L^{2,s} ({\bf R})$, 
defined for an arbitrary real $s$,  
is the weighted $L^2$-space  
with the norm 
\[
\| \psi  \|_s := \left[ \int_{{\bf R} } (1+ x^2 )^{s} 
|  \psi (x)|^2 \rmd x \right]^{1/2}  ,  
\]
where 
$\psi \in L^{2,s} ({\bf R})$ means  
$\| \psi  \|_s < \infty $. 
For positive $s$ and $s^{\prime}$, 
the relation $L^{2,s} ({\bf R}) \subset 
L^2 ({\bf R}) \subset L^{2,-s^{\prime}} ({\bf R})$ holds. 
In addition, 
we denote by ${\bf B}(s,-s^{\prime})$ the Banach space of 
the bounded operators $M$ from $L^{2,s} ({\bf R})$ 
to $L^{2,-s^{\prime}} ({\bf R})$, 
with the norm 
\[
\| M \|_{s, -s^{\prime} } := 
\sup_{\psi \in L^{2,s} ({\bf R}) ,\ \psi \neq 0} 
\frac{ \| M \psi \|_{-s^{\prime} } }{ \| \psi \|_s  } .  
\]
Notice that $M \in {\bf B}(s,-s^{\prime})$ means 
the finiteness of its norm 
$\| M \|_{s, -s^{\prime} } < \infty $. 
The reason for this kind of preparation for the free resolvent 
will be clear in the last part of the next lemma.

\begin{lm} \label{lm:R_0} 
: For $s, s^{\prime} > 1/2$, $R_0 (z) $ belongs to 
${\bf B}(s,-s^{\prime})$, and is continuously extended to 
${\overline {\bf C}}_+ \setminus \{ 0\}$ where 
${\overline {\bf C}}_+$ the closure of ${\bf C}_+$. 
\end{lm}

{\sl Proof} :  
We have an estimation 
\begin{eqnarray*}
\| R_0 (z) \psi \|_{-s^{\prime}} ^2 & = & 
\int_{{\bf R} } (1+x^2 )^{-s^{\prime}} \left| 
\frac{1}{2\rmi z^{1/2}} 
\int_{{\bf R} }  \exp(\rmi z^{1/2} |x-y|) \psi (y) \rmd y 
\right|^2 \rmd x \\ 
& \leq & \frac{\| \psi \|_s ^2 }{|2\rmi z^{1/2} |^2} 
\int_{{\bf R} } (1+x^2 )^{-s^{\prime}} \rmd x  
\int_{{\bf R} } (1+y^2 )^{-s} \rmd y 
< \infty , 
\end{eqnarray*}
for all $\psi \in L^{2,s} ({\bf R})$ and $z \in {\bf C}_+$. 
This result clearly holds for 
$z \in {\overline {\bf C}}_+ \setminus \{ 0\}$ 
and the last part of the statement is also proved.  
\qed

We  use the same symbol $R_0 (z)$ for the extension of $R_0 (z)$ 
to ${\overline {\bf C}}_+ \setminus \{ 0\}$.  
The free resolvent $R_0 (z)$ is formally expanded around $z=0$, 
\begin{equation}
R_0 (z) = \sum_{j=0} ^{\infty } (\rmi z^{1/2})^{j-1} G_j , 
\label{eqn:2.20}
\end{equation}
where $G_j$ ($j=0,1,\ldots$) is an integral operator 
acting on the suitable vectors $\psi$, 
\begin{equation}
(G_j \psi )(x):= -\frac{1}{2 j!} 
\int_{{\bf R} }  |x-y|^j \psi (y) \rmd y . 
\label{eqn:2.30}
\end{equation}

\begin{lm} \label{lm:coefficient}
: The integral operator $G_j$ 
is a Hilbert-Schmidt operator that belongs to 
${\bf B}(s,-s^{\prime})$ with $s, s^{\prime} > j+ 1/2$. 
%
\end{lm}

{\sl Proof} : Note that 
the statement in the lemma is equivalent to 
$(1+x^2 )^{-s^{\prime}/2} G_j (1+y^2 )^{-s/2}$ being 
a Hilbert-Schmidt operator on $L^2 ({\bf R})$. 
The latter is easily seen from the relation 
\[
\int_{{\bf R} } \int_{{\bf R} } 
\frac{|x-y|^{2j} }{(1+x^2 )^{s^{\prime}}  (1+y^2 )^{s}} 
\rmd x \rmd y 
\leq 2^{2j} 
\int_{{\bf R} } \int_{{\bf R} } 
\frac{|x|^{2j} + |y|^{2j} }{(1+x^2 )^{s^{\prime}}  (1+y^2 )^{s}} 
\rmd x \rmd y < \infty 
\]
for every $s, s^{\prime} > j+ 1/2$. 
\qed

The validity of the formal expansion (\ref{eqn:2.20}) is ensured 
at small energies, in the following sense.

\begin{lm} \label{lm:remainder} 
: Let $k=0,1, \ldots $.  If $R_0 (z)$ is approximated by a finite 
series in {\em (\ref{eqn:2.20})} up to $j=k$, the remainder is 
$o(|z|^{(k-1)/2})$ as $|z| \rightarrow 0$, in the norm of 
${\bf B}(s,-s^{\prime})$ with $s, s^{\prime} > k+ 1/2$. 
In the same sense, {\em (\ref{eqn:2.20})} can be differentiated in 
$z \in {\overline {\bf C}}_+ \setminus \{ 0\}$ any number of times 
for appropriate $s$ and $s^{\prime}$, that is, the $r$-th derivative 
in $z$ of the approximating finite series is equal to 
$(\rmd^r /\rmd z^r )R_0 (z)$ 
up to an error of $o(|z|^{(k-1)/2-r})$ in the norm of 
${\bf B}(s,-s^{\prime})$ with $s, s^{\prime} > k+ r+ 1/2$. 
\end{lm}

{\sl Proof} : 
We first consider the case of $k=0$. 
Suppose that $r$ is a non-negative integer, $s, s^{\prime} > r+ 1/2$, 
and $\psi \in L^{2,s} ({\bf R})$.  Then it follows that 
\begin{eqnarray}
\fl 
\left\| \frac{\rmd^r R_0 (z) }{\rmd z^r } \psi - 
\frac{\rmd^r (\rmi z^{1/2})^{-1} }{\rmd z^r } 
G_0 \psi \right\|_{-s^{\prime}} ^2 
\nonumber \\ 
\lo= 
\int_{{\bf R} } (1+x^2 )^{-s^{\prime}} 
\left| 
\int_{{\bf R} } 
\frac{\rmd^r }{\rmd z^r } 
\left[ 
\frac{1}{2\rmi z^{1/2}} 
( \exp(\rmi z^{1/2} |x-y| ) - 1 ) 
\right] 
\psi (y) \rmd y 
\right|^2 \rmd x 
\nonumber \\ 
\lo\leq  
\frac{ \| \psi \|_s ^2   }{4} 
\int_{{\bf R} } \int_{{\bf R} } 
(1+x^2 )^{-s^{\prime}}  
(1+y^2 )^{-s} 
\nonumber \\ 
\times 
\left[ 
A |z|^{-1-2r} | \exp(\rmi z^{1/2} |x-y| ) - 1 |^2 
+ \sum_{m, m^{\prime} } A_{m, m^{\prime}} |z|^m |x-y|^{m^{\prime}} 
\right] \rmd x  \rmd y   
\label{eqn:2.40}
\end{eqnarray}
where $A$ and $A_{m,m^{\prime}}$ are positive constants, 
and $m$ and $m^{\prime}$ are integers, satisfying 
$-1-2r < m \leq -1-r$ and $0 < m^{\prime} \leq 2r $, respectively. 
When $r=0$, there is no contribution from 
the summation $\sum_{m, m^{\prime} }$. 
By 
the dominated convergence theorem, 
we see that (\ref{eqn:2.40}) divided by 
$|z|^{-1-2r}$ goes to $0$ as $z \rightarrow 0$. 
%
%
In the same way, for $k=1, 2, \ldots$, we have 
\begin{eqnarray}
\fl 
\left\| \frac{\rmd^r R_0 (z) }{\rmd z^r } \psi - \sum_{j=0} ^k 
\frac{\rmd^r (\rmi z^{1/2})^{j-1} }{\rmd z^r } 
G_j \psi \right\|_{-s^{\prime}} ^2 
\nonumber \\ 
\fl= 
\int_{{\bf R} } (1+x^2 )^{-s^{\prime}}  
\nonumber \\ 
\fl~~~ \times 
\left| 
\int_{{\bf R} } 
\frac{\rmd^r }{\rmd z^r } 
\left[ 
\frac{(\rmi z^{1/2} )^{k-1} |x-y|^k }{2 (k-1)!} 
\left( 
\int_0 ^1 t^{k-1} \exp(\rmi z^{1/2} |x-y| (1-t)) \rmd t 
- \frac{1}{k} 
\right) 
\right] \psi (y) \rmd y 
\right|^2 \rmd x 
\nonumber \\ 
\fl\leq 
\frac{ \| \psi \|_s ^2   }{|2(k-1)!|^2} 
\int_{{\bf R} } 
\int_{{\bf R} } 
(1+x^2 )^{-s^{\prime}} (1+y^2 )^{-s} 
\Bigg[ 
B |z|^{k-1-2r} |x-y|^{2k} 
\nonumber \\ 
\fl~~~ \times 
\left| 
\int_0 ^1  t^{k-1} \exp(\rmi z^{1/2} |x-y| (1-t)) \rmd t 
- \frac{1}{k} 
\right|^2  
+ \sum_{m, m^{\prime} } B_{m, m^{\prime}} |z|^m |x-y|^{m^{\prime}} 
\Bigg]  \rmd x\rmd y 
\label{eqn:2.50}
\end{eqnarray}
for $s, s^{\prime} > k+ r+ 1/2$, 
where $B$ and $B_{m,m^{\prime}}$ are positive constants, 
and $m$ and $m^{\prime}$ are integers, satisfying 
$k-1-2r < m \leq k-1-r$ and $2k < m^{\prime} \leq 2(k+r) $, respectively. 
Taking the limit $z \rightarrow 0$, one can see that 
(\ref{eqn:2.50}) divided by 
$|z|^{k-1-2r}$ goes to $0$. 
This  completes the proof of the lemma. 
%
\qed

On the other hand, we also have the following lemma with respect to 
the asymptotic behaviour of $R_0(z)$ at large energies.

\begin{lm} \label{lm:large} 
: Let $k=0,1, \ldots$ and $s, s^{\prime} > k + 1/2$. 
Then $R_0 (z)$ is $k$-times differentiable in 
$z \in {\overline {\bf C}}_+ \setminus \{ 0\}$, 
in ${\bf B}(s,-s^{\prime})$, and it behaves like 
\[
\frac{\rmd^r R_0 (z) }{\rmd z^r} =O(|z|^{-(r+1)/2}) 
,~~~ r=0, 1, \ldots ,  k, 
\]
as $|z| \rightarrow \infty $ in the norm of 
${\bf B}(s,-s^{\prime})$. 
\end{lm}

{\sl Proof} : Suppose that $s, s^{\prime} > k+ 1/2$ 
and $\psi \in L^{2,s} ({\bf R})$, 
we have 
\begin{eqnarray*}
\left\| \frac{\rmd^r R_0 (z) }{\rmd z^r } \psi \right\|_{-s^{\prime}} ^2 
& & \leq 
\frac{ \| \psi \|_s ^2   }{4}
\int_{{\bf R} } \rmd x~ 
\frac{1}{(1+x^2 )^{s^{\prime}} } 
\int_{{\bf R} } \rmd y~ 
\frac{1}{(1+y^2 )^{s}} \\ 
& & ~~~ \times 
\left[ 
D |z|^{-1-r} |x-y|^{2r} 
+ \sum_{m, m^{\prime} 
  } D_{m, m^{\prime}} |z|^m |x-y|^{m^{\prime}}
\right] < \infty , 
\end{eqnarray*}
where $D$ and $D_{m,m^{\prime}}$ are positive constants, 
and $m$ and $m^{\prime}$ are integers, satisfying 
$-1-2r \leq  m < -1-r$ and $0 \leq m^{\prime} < 2r $, respectively. 
Then, the right-hand side is $O(|z|^{-1-r}) $ 
as $|z| \rightarrow \infty $. 
\qed


\section{Asymptotic expansion of the time evolution operator} 
\label{sec:6}

\setcounter{df}{0}

In order to derive the asymptotic expansion of $\exp(-\rmi tH_0) $ in (\ref{eqn:1.100}), 
we first define the spectral density denoted by 
$E^{\prime} (\lambda ) := (2\pi \rmi )^{-1} ( R_0 (\lambda ) - 
\overline{R_0 }(\lambda ) )$ for all $\lambda >0$, where   
\[
(\overline{R_0} (\lambda ) \psi ) (x) := \frac{1}{2\rmi \lambda^{1/2}} 
\int_{{\bf R} } \rmd y~ \exp(-\rmi \lambda^{1/2} |x-y| ) \psi (y) , 
\]
for every $\psi \in L^{2,s} ({\bf R})$ with $s > 1/2$.  
The operator $\overline{R_0} (\lambda ) $ 
is considered as the limit of 
$R_0 (\lambda +\rmi \epsilon )$ in $\epsilon \uparrow 0$. 
$E^{\prime} (\lambda )$ clearly belongs to 
${\bf B}(s,-s^{\prime})$ with $s, s^{\prime} > 1/2$, 
and it has the same properties as $R_0 (\lambda )$ described in 
Lemmas \ref{lm:R_0}, \ref{lm:remainder}, and \ref{lm:large}. 
%
Substituting the expansion (\ref{eqn:2.20}) and the corresponding one 
of $\overline{R_0} (\lambda ) $ into $E^{\prime} (\lambda )$, we have 
\begin{equation}
E^{\prime} (\lambda) = \pi^{-1} \sum_{j=0} ^n 
(-1)^{j-1} \lambda^{j-1/2} G_{2j} + F_n (\lambda ), 
\label{eqn:4.10}
\end{equation}
where $F_n (\lambda )$ is the remainder. 
It should be noted that there are no integer powers in $\lambda$. 
We next focus our attention on the following formula 
\begin{equation}
\exp(-\rmi tH_0) = \lim_{R \rightarrow \infty } \lim_{r \downarrow 0 } 
\int_r ^R 
E^{\prime} (\lambda ) \rme^{-\rmi t \lambda } \rmd \lambda ,  
\label{eqn:5.00}
\end{equation}
valid in ${\bf B}(s,-s^{\prime})$ for $s, s^{\prime} > 1/2$. 
The integration in the above can be also regarded as the complex integral of 
$(2\pi \rmi )^{-1} R_0 (z) \rme^{-\rmi t z }$ 
with the contour enclosing the spectrum of $H_0$, i.e., $[0, \infty )$. 
This formula is shown in Appendix\  B. 
Then, the asymptotic expansion of $\exp(-\rmi tH_0)$ at large $t$ 
is obtained from the formula (\ref{eqn:5.00}) together with 
the expansion (\ref{eqn:4.10}). To be precise, we can show  

\begin{thm} \label{thm:asymptotic evolution}
: Let $n=0, 1, \ldots$ and  $s, s^{\prime} > \max \{ 3n+3/2 , 5/2 \}$.  
Then it follows that  
\begin{equation}
\exp(-\rmi tH_0) = \pi^{-1}  \sum_{j=0} ^n 
(-1)^{j-1} \Gamma (j+1/2) (\rmi t)^{-j-1/2} ~ G_{2j} 
+ o(t^{-n-1/2}) 
\label{eqn:1.40}
\end{equation}
as $t \rightarrow \infty $, in the norm of ${\bf B}(s,-s^{\prime})$. 
\end{thm}

Note that the asymptotic form of $\exp(-\rmi tH_0)$ at large times is only determined by 
the behaviour of the free resolvent $R_0 (z)$ at small energies. 
Our proof follows the procedure proposed  
by Jensen and Kato \cite{Jensen}, and it is given in Appendix\  C.

It is interesting to rewrite the formula (\ref{eqn:1.40}) by using 
the ``generalized'' zero-energy eigenfunction of $H_0$, i.e.,  
$\varphi_0 (x) := (2 \pi )^{-1/2} $. 
Since $G_0 = - \pi \langle \varphi_0 , \cdot~ \rangle \varphi_0 $ from (\ref{eqn:2.30}), 
we have an alternative expression of (\ref{eqn:1.40}) as 
\begin{equation}
\exp(-\rmi tH_0) =  \pi^{1/2}  (\rmi t)^{-1/2} ~ 
\langle \varphi_0 , \cdot~ \rangle \varphi_0 
+ o(t^{-1/2}) , ~~~ t \rightarrow \infty . 
\label{eqn:6.40}
\end{equation}
Note that 
this has the same structure as  
the asymptotic expansions of 
the one and three dimensional systems with 
short-range potential $V$, which have no zero-energy eigenstate 
but zero-energy resonance. 
The existence of the former implies that 
there is a zero-energy eigenfunction 
belonging to the $L^2$-space, while 
that of the latter corresponds to the situation in which 
there is a function  
$\psi_0 $, that is not in $L^2 ({\bf R})$ but satisfies 
$G_0 V \psi_0 =0$ for one dimension 
or $(1+G_0^{(3)} V) \psi_0 =0$ for  three dimension. 
%
%
The system with the zero-energy resonance is known 
not to be ``generic'' \cite{Amrein,Rauch,Jensen,Murata},  
and in such a case $\varphi_0$ in (\ref{eqn:6.40}) is replaced by 
$\psi_0$. 
In this sense, the one-dimensional free particle system 
is considered to be exceptional. 
On the contrary, the three-dimensional free particle system 
seems to be generic, since the $t^{-1/2}$-term does not appear 
in the expansion of $\exp(-\rmi tH_0)$. 
This is 
because the free resolvent for the three dimensional case 
has no singularity at the origin, 
while it appears in (\ref{eqn:2.10}) (see also \cite{Muga2001}). 
To be precise, the asymptotic expansion of 
the free resolvent for the three-dimensional case is 
\[
R_0 (z) = \sum_{j=0} ^{\infty} (\rmi z^{1/2} )^j G_j^{(3)} , 
\]
where $G_j^{(3)} $ is the integral operator with the kernel 
$|{\bf x}-{\bf y}|^{j-1}/4\pi j!$ and 
${\bf x}, {\bf y}\in {\bf R}^3 $ 
\cite{Jensen}. 
Therefore the asymptote of $\exp(-\rmi tH_0)$ 
for the three-dimensional case becomes 
\begin{equation}
\exp(-\rmi tH_0) =  \pi^{3/2}  (\rmi t)^{-3/2} ~ 
\langle \varphi_0^{(3)} , \cdot~ \rangle \varphi_0^{(3)} 
+ o(t^{-3/2}) , ~~~ t \rightarrow \infty , 
\label{eqn:6.50}
\end{equation}
in ${\bf B}(s,s^{\prime })$, with large enough $s$ and $s^{\prime}$. 
Here $\varphi_0^{(3)} ({\bf x}) := (2 \pi )^{-3/2} $ 
is the zero-energy eigenfunction of $H_0$. 
Notice that $\varphi_0^{(3)} $ does not yield the $t^{-1/2}$-term 
in the expansion series, 
unlikely in the one-dimensional case.

The formula 
(\ref{eqn:1.40}) does not 
bring us the information at each point $x$. 
However, it is useful for calculating 
the quantities through the norm or the inner product, 
such as the survival probability. 
Suppose  $s, s^{\prime} > j+1/2$, 
$\sigma \in {\bf R}$, and $M \in {\bf B}(-s^{\prime}, \sigma)$. 
Then, since $G_{j }$ is considered as a vector in 
${\bf B}(s, -s^{\prime} )$, we see that 
$\|  M G_j \psi \|_{\sigma} $ 
is well defined for all $\psi \in L^{2, s} ({\bf R})$. 
Therefore, for $s, s^{\prime} > \max \{ 3n+3/2 , 5/2 \}$ and 
$M \in {\bf B}(-s^{\prime}, \sigma)$, 
we have from (\ref{eqn:1.40}) 
\begin{eqnarray}
\fl 
\left| 
\|  M \exp(-\rmi tH_0) \psi \|_{\sigma} 
- \pi^{-1}  \left\| \sum_{j=0} ^n 
(-1)^{j-1 } 
\Gamma (j+1/2) 
~ (\rmi t)^{-j-1/2} ~ M G_{2j} \psi \right\|_{\sigma} 
\right| 
\| \psi \|_{s}^{-1}  \| M \|_{-s^{\prime} ,\sigma }^{-1}  
\nonumber \\ 
\fl\leq 
\left\| 
\exp(-\rmi tH_0) - \pi^{-1}  \sum_{j=0} ^n 
(-1)^{j-1} \Gamma (j+1/2) (\rmi t)^{-j-1/2} ~ G_{2j} 
\right\|_{s,-s^{\prime} }  = o(t^{-n-1/2}) , 
%
\label{eqn:6.70}
\end{eqnarray}
for all $\psi \in L^{2,s} ({\bf R})$. 
For example, we can take $(1+x^2)^{s^{\prime} /2} E(B)$ for $M$ 
with $\sigma = -s^{\prime}$, where $E(B)$ is 
the spectral measure 
of the position operator and $B$  an arbitrary bounded interval 
of ${\bf R}$, i.e. $(E(B) \psi )(x) = \psi (x)$ ($x \in B$) or 
$0$ ($x \notin B$).  
Then, 
$\| M \exp(-\rmi tH_0) \psi \|_{-s^{\prime}}^2 = 
\| E(B) \exp(-\rmi tH_0) \psi \|^2 = \int_{ B} |\psi (x,t)|^2 \rmd x $, 
and the last quantity is called the nonescape probability, 
which is the probability to find the particle in $B$ at a time $t$. 
From a similar argument, we also have, 
\begin{eqnarray}
\fl 
\left| 
\langle \phi, \exp(-\rmi tH_0) \psi \rangle - 
\pi^{-1}  \sum_{j=0} ^n 
(-1)^{j-1} \Gamma (j+1/2) (\rmi t)^{-j-1/2} ~ 
\langle \phi, G_{2j} \psi \rangle 
\right| 
\| \phi \|_{s}^{-1}   \| \psi \|_{s}^{-1} 
\nonumber \\ 
\fl\leq 
\left\| 
\exp(-\rmi tH_0) - \pi^{-1}  \sum_{j=0} ^n 
(-1)^{j-1} \Gamma (j+1/2) (\rmi t)^{-j-1/2} ~ G_{2j} 
\right\|_{s,-s^{\prime} } 
= o(t^{-n-1/2}) , 
\label{eqn:6.80}
\end{eqnarray}
for $s, s^{\prime} > \max \{ 3n+3/2 , 5/2 \}$ with $s \geq s^{\prime}$ 
and all $\phi, \psi \in L^{2,s} ({\bf R})$. 
The asymptotic formula for the survival amplitude of $\psi$ is the special case 
of (\ref{eqn:6.80}). 
%


\section{Dependence on the initial momentum distribution 
} 
\label{sec:7}

\setcounter{df}{0}

In practical situations, it sometimes happens that 
there are no contributions from some 
of the $G_{2j}$'s 
to such quantities like 
$\| M \exp(-\rmi tH_0) \psi \|_{\sigma} $ or 
$\langle  \phi,  \exp(-\rmi tH_0) \psi \rangle $, 
when they act on a certain  vector $\psi$. 
In this section, we confine ourselves to 
such situations for  
the survival amplitude $\langle  \psi,  \exp(-\rmi tH_0) \psi \rangle $. 

\begin{lm} \label{lm:moments} 
: Let $n=0,1, \ldots$. 
If $s > n + 1/2$ and $\psi \in L^{2,s} ({\bf R})$, 
then $x^j \psi (x) \in L^1 ({\bf R})$, for all $j=0,1,\ldots, n$. 
\end{lm}

{\sl Proof} :  The statement is obtained straightforwardly: 
$ 
\int_{\bf R} |x^j \psi (x) | \rmd x = 
\int_{\bf R} (1+x^2 )^{-(s-j)/2} (1+x^2 )^{(s-j)/2} 
|x^j \psi (x) | \rmd x \leq 
[ \int_{\bf R} (1+x^2 )^{-(s-j)} \rmd x ]^{1/2} 
[ \int_{\bf R} (1+x^2 )^s |\psi (x) |^2 \rmd x ]^{1/2} < \infty  
$
for all $j=0,1,\ldots, n$ with $s > n + 1/2$. 
\qed

\begin{lm} \label{lm:w-momentum}
: Let $m=1,2, \ldots$. If $s  > \max \{ 2(m-1), m \} + 1/2 $ 
and 
$\psi \in L^{2,s} ({\bf R})$, then the following three statements 
are equivalent: 

\noindent 
~ (a)~  $\hat{\psi} (k) = O(k^{m})$,  $k \rightarrow 0$. \\ 
~ (b)~  $\langle \psi , G_{2j} \psi \rangle = 0, ~ j=0,1,\ldots, m-1$. \\
~ (c)~  $\int_{\bf R} x^j \psi (x) \rmd x = 0, ~ j=0,1,\ldots, m-1$. 

\noindent
In particular, 
we have 
$G_{2j} \psi$ 
$ =0 , ~ j=0,1,\ldots, (m-1)/2$ for odd $m$, 
or $G_{2j} \psi =0 , ~ j=0,1,\ldots, m/2-1$ for even $m$. 
\end{lm}


{\sl Proof} :  
Suppose that $\psi \in L^{2,s} ({\bf R})$ with $s  
>  \max \{ 2(m-1), m \} + 1/2 $. Then by Lemma \ref{lm:moments}, 
we first see that 
$x^j \psi (x) \in L^1 ({\bf R})$ for 
$j=0,1,\ldots,  \max \{ 2(m-1), m \} $. 
Since 
$
\rme^{-\rmi kx} = \sum_{j=0} ^{m-1} (-\rmi kx)^j /j! + (-\rmi kx)^{m} 
\int_{0} ^{1} t^{m-1} \rme^{-\rmi kx(1-t)} \rmd t /(m-1)! 
$,  
the fact that (c) implies (a) immediately follows, by using   
\begin{eqnarray}
\fl 
|\hat{\psi} (k)| = 
\left| 
\frac{1}{\sqrt{2\pi }} \int_{\bf R} 
\left[ 
\sum_{j=0} ^{m-1} \frac{(-\rmi kx)^j }{j! } + 
\frac{(-\rmi kx)^{m} }{(m-1)!} 
\int_{0} ^{1} t^{m-1} \rme^{-\rmi kx(1-t)} \rmd t 
\right] \psi (x) \rmd x 
\right| 
\label{eqn:7.10} \\ 
\lo\leq  
\frac{|k|^{m} }{\sqrt{2\pi } (m-1)!} \int_{\bf R} 
|x^{m} \psi (x) | \rmd x = O(k^m ),  ~~ k \rightarrow 0. 
\nonumber 
\end{eqnarray}
To prove the fact that (a) implies (c), let us remember that 
$\hat{\psi} (k) = O(k^{m})$ ($k \rightarrow 0$) means that 
for some $\delta$ and some finite $C \geq 0$, 
$|\hat{\psi} (k) / k^m| \leq C$ for all $k$ satisfying 
$0< |k| < \delta$. 
Then, from (\ref{eqn:7.10}),  
$\int_{\bf R}  x^j \psi (x) \rmd x $ should vanish 
for all $j=0,1,\ldots, m$. 
The fact that (c) implies (b) follows straightforwardly 
from the identity, 
\begin{equation}
-2 (2j)! \langle \psi , G_{2j} \psi \rangle  = 
\sum_{i=0} ^{2j} (-1)^{2j-i}  {2j \choose i } 
\int_{\bf R} x^{2j-i} \overline{ \psi (x) } \rmd x 
\int_{\bf R} y^i \psi (y) \rmd y, 
\label{eqn:7.20}
\end{equation}
where the bar ( $\bar{ }$ ) denotes the complex conjugate. 
To prove the fact that (b) implies (c), 
we first use the assumption 
that $-2 \langle \psi , G_{0} \psi \rangle  = 
| \int_{\bf R} \psi (x) \rmd x |^2 =0$. Then from (\ref{eqn:7.20}) 
for $j=1$ we have $| \int_{\bf R} x \psi  (x) \rmd x |^2 =0$, 
from which the remaining equalities recursively follow. 
For the proof of the last part of the lemma, 
we note that for $s, s^{\prime}  >  2j + 1/2 $ and  
$\psi \in L^{2,s} ({\bf R})$,  
\[
G_{2j} \psi =0 ~~ \Leftrightarrow ~~ 
\| G_{2j} \psi \|_{-s^{\prime }}  = 0  ~~ \Leftrightarrow ~~ 
\int_{\bf R} x^i \psi (x) \rmd x =0 ~(i=0,1,\ldots ,2j). 
\]
Hence, if (c) holds, we obtain the equality  
$G_{2j} \psi =0$ for all $j$ satisfying $2j \leq m-1$. 
This completes the proof.  
\qed

Now we shall derive 
the asymptotic formula for the survival amplitude 
by combining Theorem\ \ref{thm:asymptotic evolution} with 
Lemma\ \ref{lm:w-momentum}. 
The asymptotic formula itself 
immediately follows from (\ref{eqn:6.80}) with $\phi = \psi $,  
under the assumption in Theorem\ \ref{thm:asymptotic evolution}. 
We also see that 
$\max \{ 3n+3/2 , 5/2 \} \geq 3n +3/2 \geq 
\max \{ 2(m-1), m \} + 1/2$ for $m \leq n$, and thus 
the assumption in 
Lemma\ \ref{lm:w-momentum} is included in that in 
Theorem\ \ref{thm:asymptotic evolution}. 
Hence, we finally obtain 
the following theorem for the survival amplitude of $\psi$, 
which is closely connected to 
the behaviour of $\psi$ at zero momentum.

\begin{thm} \label{thm:dep on momentum}
: Let $m, n, m \leq n$ be non-negative integers, 
$s 
> \max \{ 3n+3/2 , 5/2 \}$, and 
$\psi \in L^{2,s} ({\bf R})$.  
Then it follows that 
\begin{equation}
\fl 
\langle \psi, \exp(-\rmi tH_0) \psi \rangle = 
\pi^{-1}  \sum_{j=0} ^n 
(-1)^{j-1} \Gamma (j+1/2) (\rmi t)^{-j-1/2} ~ 
\langle \psi, G_{2j} \psi \rangle + o(t^{-n-1/2}) .  
\label{eqn:1.50}
\end{equation}
In particular, for some $m \geq 1$, if 
$\hat{\psi} (k) = O(k^{m})$ as $k \rightarrow 0$, then 
$\langle \psi, G_{2j} \psi \rangle =0$ for all $j=0,1, \ldots , m-1,$ 
and vice versa. 
\end{thm}

The asymptotic formula (\ref{eqn:1.50}) can also be written as the form 
without use of $G_{2j}$'s. 
Let $n=0,1,\ldots$. If $s> n+1/2$ and $\psi \in L^{2,s} ({\bf R})$, 
$\hat{\psi} (k)$ is $n$-times continuously differentiable in $k$  
with 
$
\hat{\psi}^{(n)} (k)  = (-\rmi )^n (2\pi )^{-1/2} 
\int_{\bf R} x^n \rme^{-\rmi kx} \psi (x) \rmd x. 
$
Then, we have 
from Lemma\ \ref{lm:w-momentum} and (\ref{eqn:7.20}) 
that 
if $\hat{\psi} (k) = O(k^{m})$ as $k \rightarrow 0$, 
\begin{equation}
 \langle \psi, G_{2m} \psi \rangle  
 = 
- \frac{(-1)^{m} }{2 (2m)! } {2m \choose m } 
\left| \int_{\bf R} x^{m} \psi (x)  \rmd x \right|^2 
= 
- \frac{(-1)^{m} \pi }{(m!)^2 } 
|\hat{\psi}^{(m)} (0) |^2 . 
\label{eqn:7.40}
\end{equation}
This expression also holds for $m=0$ [see (\ref{eqn:6.40})]. 
Therefore, we obtain from (\ref{eqn:1.50}) 
the asymptotic formula for 
the survival probability 
\begin{equation}
| \langle \psi, \exp(-\rmi tH_0) \psi \rangle |^2 = 
t^{-2m-1} \frac{\Gamma (m+1/2)^2 }{(m!)^4} 
|\hat{\psi}^{(m) } (0)|^4 + o(t^{-2m-1} ) .  
\label{eqn:7.50}
\end{equation}
In Theorem\ \ref{thm:dep on momentum}, 
the assumption that $\psi \in L^{2,s} ({\bf R})$ 
with sufficiently large $s$ is technically required. 
According to the expression in (\ref{eqn:7.50}), 
it is worth reviewing this assumption in the momentum representation. 
Let $\psi \in L^{2,s} ({\bf R})$ ($s\geq 0$) and 
$[s]$ denote the smallest integer less than or equal to $s$. 
Then, by the Plancherel theorem, we have 
\[
\infty > \int_{\bf R} (1+x^2 )^n |\psi (x)|^2 \rmd x 
\geq \int_{\bf R}  |x^n \psi (x)|^2 \rmd x 
= \int_{{\bf R}_k }  |{\hat{ \psi }}^{(n)} (k)|^2 \rmd k 
\] 
for all $n=0,1, \ldots [s].$ 
It should be noted here that $\hat{\psi} (k) $ is implicitly 
guaranteed to be $[s]$-times differentiable. 
Hence, as an obvious case, we can find the following subspace 
\begin{equation}
{\cal D}:= \{ \hat{\psi} \in C^{\infty} ({\bf R}_k ) ~|~ 
\hat{\psi}^{(n)} \in L^2 ({\bf R}_k ) ,~ n=0,1,\ldots \} , 
\label{eqn:7.60}
\end{equation}
which satisfies ${\cal D} \subset 
\{ \hat{\psi} \in L^2 ({\bf R}_k ) ~|~ 
\psi \in L^{2,s} ({\bf R}) \}   
$
for all $s \geq 0$.  
Then, Equations (\ref{eqn:6.70}) and (\ref{eqn:6.80})  
with an arbitrary $n$ can be applied to the wave functions 
belonging to the above ${\cal D}$.    
Examples of such (initial) wave functions include  
$ k^l/(1+ k^{2m})^{\alpha } $ and 
$k^n \exp(-k^2)$, where $l,m,n =0,1, \ldots$ and  
$\alpha > 0$ with $2m\alpha -l > 1/2$.

In order to see some implications of Theorem\ \ref{thm:dep on momentum}, 
let us refer to the following two examples. 
We first consider the rapidly decreasing functions, 
$\hat{\phi_m } (k) = N_m k^m \exp(-a_0 k^2)$, as initial wave functions, 
where  
$m =0,1,\ldots$, $a_0 >0$, and $N_{m} := [\Gamma (m+1/2)/
(2a_0 )^{m+1/2} ]^{-1/2}$ 
being the normalization constants. 
Then it is obtained through the Laplace transform 
\begin{eqnarray}
\fl 
\langle \phi_m , \exp(-\rmi tH_0) \phi_m \rangle  
& = & 
\int_{-\infty } ^{\infty} |\hat{\phi_m } (k) |^2 \exp(-\rmi tk^2 ) \rmd k 
\nonumber \\ 
& = & [1+\rmi t /(2 a_0)]^{-(m + 1/2)} 
= (\rmi t/2a_0)^{-(m+1/2)} [1+ O(t^{-1})]. 
\label{eqn:7.70}
\end{eqnarray}
On the other hand, we see  
that the right-hand side of (\ref{eqn:7.50}) 
exactly corresponds to that in 
(\ref{eqn:7.70}) in the leading order. 
The other example is a special case that $\hat{\psi}^{(n)} (0) =0 $ 
for all $n=0,1, \ldots$. 
It is worth noticing that for such a initial wave function 
we clearly see from (\ref{eqn:7.50}) that 
\[
| \langle \psi, \exp(-\rmi tH_0) \psi \rangle |^2 = 
 o(t^{-2n-1} ) , 
\]
for every $n \geq 0$. That is, 
the survival probability decays faster than any power of $t^{-1}$. 
However, it must decay slower than any exponential at long times 
for $H_0 \geq 0$ \cite{Fonda}. 
This strange decay behaviour is also found in 
a study of the time operator (Proposition 3.2 in \cite{Miyamoto}). 
The set of such a special wave function is given, e.g., by 
$
{\cal C}_{\rm i}:= \{ \hat{\psi} \in C^{\infty}_0 ({\bf R}_k ) ~|~ 
\exists k_0 >0;  \hat{\psi} (k) =0, 
\mbox{ for } k \in [-k_0 , k_0 ]  \}  
$. 
We see from (\ref{eqn:7.60}) 
that ${\cal C}_{\rm i} \subset {\cal D}$. 
A wave function in ${\cal C}_{\rm i}$ 
has a positive lower-bound $k_0^2$ on energy. 
For instance, the following function 
\[
\hat{\psi} (k) = 
\left\{ 
\begin{array}{cc} 
\exp(-1/[k_0^2 -(k-d)^2 ]) & (|k-d|< k_0 ) \\ 
0 & (|k-d| \geq k_0 ) 
\end{array}
\right. , 
\]
where $d> k_0 > 0$, surely belongs to  ${\cal C}_{\rm i}$.


\section{Concluding remarks} 
\label{sec:8}

\setcounter{df}{0}

We have derived the asymptotic expansion of the time evolution 
operator for the one-dimensional free particle system, 
in terms of the operators which are 
expansion coefficients of the free resolvent at small energies.  
This enables us to obtain the asymptotic formula for the survival 
probability of $\psi$, and also to evaluate, 
in a systematic way, 
the condition for 
the initial wave function $\psi$ which makes 
the first several terms of the asymptotic formula vanish.  
%
%
We have found that if $\hat{\psi} (k) =O(k^m)$ for some 
non-negative integer $m$ at zero momentum, 
the asymptotic power of $t^{-1}$ for the survival probability 
must be $2m+1$. 
In other words, the information about 
the initial momentum distribution in the vicinity of zero momentum 
is reflected in the asymptotic decay form $t^{-2m-1}$ at long times. 
Our results are essentially due to the choice of 
the initial wave functions $\psi$ in $L^{2,s} ({\bf R})$ 
with sufficiently large $s$. 
This guarantees the existence of 
the higher derivatives at zero momentum 
(see Remark \ref{momentum representation}). 
However, there is another wave function such that  
${\hat{\psi}}^{(n)} (0) =0$ up to $n=m-1$, 
while its $m$-th derivative ${\hat{\psi}}^{(m)} (0)$  diverges. 
Related wave functions are considered in 
\cite{Unnikrishnan,Lillo,Mendes,Damborenea}. 
For such states, the asymptotic formula (\ref{eqn:7.50}) 
is not correct. Indeed, the actual asymptotic decay form 
of the survival probability includes 
terms of non-odd power of $t^{-1}$. 
We hope to address this issue in the future. 
%



\section*{Acknowledgements} 

The author would like to thank Professor I.\ Ohba 
and Professor H.\ Nakazato 
for useful and helpful  discussions. 


\appendix 

\section*{Appendix\ A} 

\label{sec:a}

\setcounter{df}{0}

\makeatletter
 \renewcommand{\theequation}{%
       A\arabic{equation}}
 \@addtoreset{equation}{section}
\makeatother

Dependence of the survival probability on the initial state $\psi$ 
for one-dimensional free particle system 
can be seen in the following inequality 
(Theorem\ 4.1 and section VI in \cite{Miyamoto}), 
\begin{equation}
 | \langle \psi , \exp(-\rmi tH_0 ) \psi   \rangle |^2 
 \leq 
 \frac{4 \| T_0  \psi \|^2  \| \psi \|^2 }{t^2} , 
~~ t \in {\bf R}, 
 \label{eqn:1.20}
\end{equation} 
where 
$T_0$ is the Aharonov-Bohm time operator \cite{Aharonov}. 
This brings us with an interpretation of $\| T_0  \psi \|$, 
or of the time uncertainty calculated from $T_0$.  
The relevant information herein is that for 
any $L^2$-function $\psi $ whose several moments are finite, 
e.g., $\psi \in L^{2,s} ({\bf R})$ ($s>3/2$),  
we have 
\begin{equation}
\| T_0  \psi \| < \infty ~~ \Longleftrightarrow ~~ 
\hat{\psi}(k) = O(k^2 ), ~~ k \rightarrow 0. 
 \label{eqn:1.30}
\end{equation} 
This implies, together with (\ref{eqn:1.20}), that 
the condition (\ref{eqn:1.30}) at zero momentum imposes  on  
the survival amplitude 
$ \langle \psi , \exp(-\rmi tH_0 ) \psi \rangle  $ 
a decay faster than $t^{-1}$, 
which is obviously faster than $t^{-1/2}$ of the usual decay law 
for the one dimensional free particle system. 

To prove the relation\ (\ref{eqn:1.30}), 
we first define the Aharonov-Bohm time operator $T_0$, which is mathematically 
well treated in the scheme of the axiomatic quantum mechanics 
\cite{Miyamoto,Egusuquiza}. 
We define this operator as follows: the domain of $T_0$ is 
\begin{equation} 
\fl 
D(T_0 ) := 
              \left\{ 
              \psi \in L^2 ({\bf R}) ~\left| ~
              \lim_{k \rightarrow 0} 
              \hat{ \psi} (k) / |k|^{1/2} =0 
              \mbox{ and } 
              (\widehat{T_0 \psi})(k)  \in L^2 ({\bf R}_k )
              \right. 
              \right\} , 
              \label{eqn:a10}
\end{equation}
and its action  
\begin{equation}
\fl 
  (\widehat{T_0 \psi})(k)  = \frac{\rmi }{4} \left( 
  \frac{\rmd \hat{ \psi} (k) /k}{\rmd k} + \frac{1}{k}
  \frac{\rmd \hat{ \psi} (k)}{\rmd k}  \right) ,~~~
  \mbox{\rm a.e. } k \in {\bf R}_k   , ~~~ 
  \psi \in D( T_0 ), 
\label{eqn:a20}
\end{equation}
where $\hat{ \psi} (k)$ is assumed to be differentiable everywhere 
except the origin. 
For a $\psi$ that belongs to $L^{2,s} ({\bf R})$ ($s>3/2$), 
$\psi (x)$ and $x \psi (x)$ in $L^1 ({\bf R})$ by Lemma\ \ref{lm:moments}. 
This implies $\hat{ \psi} (k)$ to be differentiable everywhere including the origin. 
Then, 
it follows as in (\ref{eqn:7.10}) that for $k \rightarrow 0$ 
\begin{equation}
\hat{\psi} (k)  =  
\hat{\psi} (0)
+ k  \hat{\psi}^{\prime } (0)
+ O(k^2), ~~~ 
\hat{\psi}^{\prime} (k) = \hat{\psi}^{\prime} (0) + O(k). 
\label{eqn:a30}
\end{equation}
Hence 
\begin{equation}
  (\widehat{T_0 \psi})(k)  = \frac{\rmi }{4} \left( 
  - \frac{\hat{ \psi} (0) }{k^2} 
  + \frac{\hat{ \psi}^{\prime} (0)}{k}  \right) + O(1) 
\label{eqn:a50}
\end{equation}
for small $k$. 
Furthermore, 
since $\psi \in L^{2,s} ({\bf R}) \subset L^{2,1} ({\bf R}) $, 
$\int_{\bf R} |x\psi (x) |^2 \rmd x = \int_{{\bf R}_k} 
|{\hat{\psi}}^{\prime} (k) |^2 \rmd k < \infty  $.  
Thus, 
$(\widehat{T_0 \psi})(k) $ is assured to be square integrable on 
$(-\infty, -\delta ] \cup [\delta, \infty)$ 
for an arbitrary $\delta >0$. 
%

Now, to prove the fact that $\| T_0  \psi \| < \infty $ implies 
$\hat{\psi}(k) = O(k^2 )$, $k \rightarrow 0$ in the relation\ (\ref{eqn:1.30}), 
let us suppose that $(\widehat{T_0 \psi})(k) \in L^2 ({\bf R}_k )$. 
Then, from the general property of $L^2$-functions, 
we see that 
$(\widehat{T_0 \psi})(k) \in L^1 ([-\delta, \delta ] )$. 
This contradicts with (\ref{eqn:a50}) unless 
$\hat{\psi} (0) = \hat{\psi}^{\prime } (0) =0$. 
Thus we have that $\hat{\psi} (k) = O(k^2 )$. Conversely, 
if $\hat{\psi} (k) = O(k^2 )$, it follows from (\ref{eqn:a30}) 
that $\hat{\psi} (0) = \hat{\psi}^{\prime } (0) =0$. 
Then, (\ref{eqn:a50}) implies that 
$(\widehat{T_0 \psi})(k) \in L^2 ([-\delta, \delta ] )$. 
Hence, $(\widehat{T_0 \psi})(k)$ belongs to $L^2 ({\bf R}_k )$, 
and the proof of the relation\ (\ref{eqn:1.30}) is completed.


\section*{Appendix\ B} 

\label{sec:b}

\setcounter{df}{0}

\makeatletter
 \renewcommand{\theequation}{%
       B\arabic{equation}}
 \@addtoreset{equation}{section}
\makeatother

\setcounter{equation}{0}

In this appendix, we derive the formula\ (\ref{eqn:5.00}) 
which directly relates $\exp(-\rmi tH_0)$ to the spectral density. 
Let us remember that 
the time evolution operator for one dimensional free particle 
system is explicitly represented as 
\begin{equation}
\fl 
\psi (x,t) = (\exp(-\rmi t H_0) \psi )(x) = (4\pi \rmi t)^{-1/2}
\int_{\bf R} \exp(\rmi |x-y|^2 /4t) 
\psi (y)\ \rmd y =  O(t^{-1/2}),   
\label{eqn:1.10}
\end{equation}
for all $\psi \in L^1 ({\bf R} ) \cap L^2 ({\bf R} )$ 
and $t \neq 0$ \cite{RS2-1}. 
Note that $L^{2,s} ({\bf R} ) \subset L^1 ({\bf R} )$ 
for $s>1/2$.  
Then, $\exp(-\rmi tH_0)$ is considered as an integral operator 
belonging to ${\bf B}(s,-s^{\prime})$ for $s, s^{\prime} > 1/2$. 

Let $s, s^{\prime} > 1/2$ and  $\psi \in L^{2,s} ({\bf R} )$. 
We have an equality 
\[
\fl 
\left( \int_r ^R 
E^{\prime} (\lambda ) \rme^{-\rmi t \lambda } \rmd \lambda \psi \right) (x) 
= 
\int_{{\bf R} } \rmd y~ 
\left[ 
\int_r ^R 
\frac{\cos (\lambda^{1/2} |x-y|) \rme^{-\rmi t \lambda }}{2\pi \lambda^{1/2} } 
\rmd \lambda 
\right]  \psi (y) ,~~~ {\rm a.e. }~ x \in {\bf R} , 
\]
for positive $r$ and $R$. Then it follows from (\ref{eqn:1.10}) that 
\begin{eqnarray}
\fl 
\left\| 
\int_r ^R E^{\prime} (\lambda ) \rme^{-\rmi t \lambda } \rmd \lambda \psi 
- \exp(-\rmi tH_0) \psi 
\right\|_{-s^{\prime}} ^2 
\| \psi \|_s ^{-2} 
\nonumber \\ 
\fl\leq  
\int_{{\bf R} } \int_{{\bf R} } 
\left| 
\int_r ^R 
\frac{\cos (\lambda^{1/2} |x-y|) \rme^{-\rmi t \lambda }}{2\pi \lambda^{1/2} } 
\rmd \lambda 
- \frac{\exp(-|x-y|^2 /4\rmi t )}{(4\pi \rmi t)^{1/2} } 
\right|^2 
\frac{\rmd x\rmd y  }{(1+x^2 )^{s^{\prime}} (1+y^2 )^{s} } . 
\label{eqn:5.10}
\end{eqnarray}
We here see that 
\begin{equation}
\sup_{x\in {\bf R} , R>0} \left| 
\int_0 ^R 
\frac{\cos (\lambda^{1/2} |x|) \rme^{-\rmi t \lambda }}{2\pi \lambda^{1/2} } 
\rmd \lambda 
\right| < \infty . 
\label{eqn:5.20}
\end{equation}
To derive this, we have assumed  $t>0$, however, the following argument  
can be applied to negative $t$,  
\begin{eqnarray*}
\fl
\left| 
\int_0 ^R 
\frac{\cos (\lambda^{1/2} |x|) \rme^{-\rmi t \lambda }}{2\pi \lambda^{1/2} } 
\rmd \lambda   
\right| 
& & = 
(2\pi)^{-1} 
\left| 
\int_{-R^{1/2}} ^{R^{1/2}} \exp(-\rmi t(\xi + a)^2 ) \rmd \xi 
\right| \\ 
& & = 
(2\pi)^{-1} 
\left| 
\left[ 
\int_0 ^{R^{1/2} +a} \exp(-\rmi t\xi^2 ) \rmd \xi 
+ \int_{-R^{1/2} +a} ^0 \exp(-\rmi t\xi^2 ) \rmd \xi \right] 
\right| , 
\end{eqnarray*}
where $a=|x|/2t$. 
The last two integrals are estimated, by the complex integrals 
with the contour in fourth quadrant, to be  
\begin{eqnarray*}
\fl 
\left| 
\int_0 ^{|R^{1/2} \pm a|} \exp(-\rmi t\xi^2 ) \rmd \xi 
\right| 
&\leq& 
\frac{\pi (1- \exp(-t |R^{1/2} \pm a|^2 ) ) }{4t |R^{1/2} \pm a|} 
+ \int_0 ^{|R^{1/2} \pm a|} \exp(-tr^2 ) \rmd r 
\\
&\leq& 
\frac{\pi}{4t} \sup_{x>0} \frac{1-\exp(-tx^2) }{x} 
+\frac{1}{2} \sqrt{\frac{\pi}{t}} < \infty . 
\end{eqnarray*}
This leads to (\ref{eqn:5.20}). 
We also see straightforwardly 
\begin{equation}
\lim_{R \rightarrow \infty } \lim_{r \downarrow 0 } 
\int_r ^R 
\frac{\cos (\lambda^{1/2} |x-y|) \rme^{-\rmi t \lambda }}{2\pi \lambda^{1/2} } 
\rmd \lambda 
= \frac{\exp(-|x-y|^2 /4\rmi t)}{(4\pi \rmi t)^{1/2} } . 
\label{eqn:5.30}
\end{equation}
Therefore, 
by the dominated convergence theorem, 
we have from (\ref{eqn:5.10}), (\ref{eqn:5.20}), and (\ref{eqn:5.30}) 
\[
\sup_{\psi \in L^{2,s} ({\bf R}) , \psi \neq 0} 
\left\| \exp(-\rmi tH_0) \psi - 
\int_r ^R E^{\prime} (\lambda ) \rme^{-\rmi t \lambda } \rmd \lambda \psi 
\right\|_{-s^{\prime}} ^2 \| \psi \|_s ^{-2} 
\rightarrow 0  
\]
as $r \downarrow 0$ and  $R \rightarrow \infty $, 
and we finally obtain the formula\ (\ref{eqn:5.00}). 


\section*{Appendix\ C} 

\label{sec:c}

\setcounter{df}{0}

\makeatletter
 \renewcommand{\theequation}{%
       C\arabic{equation}}
 \@addtoreset{equation}{section}
\makeatother

\setcounter{equation}{0}

In order to prove Theorem\ \ref{thm:asymptotic evolution}, 
we first summarize the several property of $E^{\prime} (\lambda)$. 
By Lemma \ref{lm:remainder}, we see that for $n\geq 0$ 
the remainder $F_n (\lambda )$ in (\ref{eqn:4.10}) is ($n+1$)-times differentiable, 
in ${\bf B}(s,-s^{\prime})$ 
for $s, s^{\prime} > 2n+(n+1)+1/2 = 3n+3/2 $, 
and satisfies that  
$
(\rmd / \rmd \lambda )^r F_n (\lambda )  = o(\lambda^{n -r -1/2} ) 
$
as $\lambda \downarrow 0$ ($r=0, 1, \ldots ,  n+1$). 
On the other hand, we see from Lemma \ref{lm:large} that  
$ (\rmd / \rmd \lambda )^r E^{\prime} (\lambda ) = O(\lambda^{-(r+1)/2})$ 
as $\lambda \rightarrow \infty $, in ${\bf B}(s,-s^{\prime})$ for 
$s, s^{\prime} > (m+1)+1/2 = m+3/2$ ($ r=0,1, \ldots , m+1$).  
In particular, if $m \geq 1$,  
$(\rmd /\rmd \lambda )^{m+1} E^{\prime} (\lambda )$ is integrable 
on $[\delta , \infty )$ for an arbitrary $\delta >0$.

Let us now split the integral in (\ref{eqn:5.00}) 
into two parts by writing 
\[
E^{\prime} (\lambda ) = \phi (\lambda ) E^{\prime} (\lambda ) 
+ (1- \phi (\lambda ) ) E^{\prime} (\lambda ) , 
\]
where $\phi \in C_0 ^{\infty} ([0,\infty ) )$ and satisfies  
$\phi (\lambda ) =1$ in a neighbourhood of $\lambda = 0$. 
Such a function is realized by $f (\lambda ) = 1- \int_0 ^\lambda 
g(x) \rmd x$, where $g(x) = h(x)/\int_{\bf R} h(x) \rmd x$ and 
$h(x) = \exp(-1/[1-(x-d)^2 ])$ ~($|x-d|<1$) or $0$ ($|x-d| \geq 1$) 
with $d>1 $.

From Lemma 10.1 in \cite{Jensen} and the discussion as mentioned above, we see that 
$(1- \phi (\lambda ) ) E^{\prime} (\lambda ) $ has a contributions of 
$o(t^{-m-1})$ to $\exp(-\rmi tH_0)$ in ${\bf B}(s,-s^{\prime})$, 
where $m \geq 1$ and  $s, s^{\prime} > m+3/2$.

On the other hand, the contribution of 
$\phi (\lambda ) E^{\prime} (\lambda ) $ to $\exp(-\rmi tH_0)$ 
gives 
the main part of the asymptotic expansion. 
Then, the coefficient of $G_{2j} $ is given by 
\begin{eqnarray}
\fl 
\int_0 ^{\infty } 
\phi (\lambda ) \lambda^{j-1/2} \rme^{-\rmi t \lambda } ~ \rmd \lambda 
& = & 
i^j \frac{\rmd^j}{\rmd t^j } 
\left[ 
\int_0 ^{\infty } \lambda^{-1/2} \rme^{-\rmi t \lambda } ~ \rmd \lambda 
+ 
\int_0 ^{\infty } (\phi (\lambda ) -1) 
\lambda^{-1/2} \rme^{-\rmi t \lambda } ~ \rmd \lambda 
\right]  \nonumber \\ 
& = & 
\Gamma (j+1/2)  (\rmi t)^{-j-1/2} 
+ 
i^j \frac{\rmd^j}{\rmd t^j } \int_0 ^{\infty } (\phi (\lambda ) -1) 
\lambda^{-1/2} \rme^{-\rmi t \lambda } ~ \rmd \lambda . 
\label{eqn:6.10}
\end{eqnarray}
Note that 
since $\phi (\lambda ) -1$ and all its  
derivatives vanish in the neighbourhood of $\lambda =0$ and 
$\phi \in C_0 ^{\infty} ([0,\infty ) )$, 
the last term in (\ref{eqn:6.10}) 
decays faster than any 
negative-power of $t$. 
Furthermore, we understand, from Lemma\ 10.2 in \cite{Jensen} 
and the discussion in the first of the appendix, that 
if $s, s^{\prime} > 3n+3/2 $, any contribution of 
the Fourier transform of the remainder $\phi (\lambda ) F_n (\lambda ) $ 
to $\exp(-\rmi tH_0)$ is of $o(t^{-n-1/2})$ in 
the norm of ${\bf B}(s,-s^{\prime})$. 
%
Summarizing the above arguments, we finally obtain,   
under the condition $s, s^{\prime } > \max \{ 3n+3/2, 5/2\}$,  
\begin{eqnarray*}
\fl
\left\| \exp(-\rmi tH_0) - \pi^{-1}  \sum_{j=0} ^n 
(-1)^{j-1} \Gamma (j+1/2) (\rmi t)^{-j-1/2} ~ G_{2j} 
\right\|_{s,-s^{\prime} }  \\ 
\fl\leq 
\left\| 
\exp(-\rmi tH_0) - \int_r ^R 
E^{\prime} (\lambda ) \rme^{-\rmi t \lambda } \rmd \lambda 
\right\|_{s,-s^{\prime} }    \\   
\fl~~~ 
+ 
\left\| 
\int_r ^R 
E^{\prime} (\lambda ) \rme^{-\rmi t \lambda } \rmd \lambda 
- \sum_{j=0} ^n 
(-1)^{j-1} \Gamma (j+1/2) (\rmi t)^{-j-1/2} ~ G_{2j} 
\right\|_{s,-s^{\prime} }   \rightarrow 0 + o(t^{-n-1/2}) ,  
\end{eqnarray*}
as $r \downarrow 0 $ and $R \rightarrow \infty$.  
This is just the asymptotic expansion of $\exp(-\rmi tH_0)$ in (\ref{eqn:1.40}).

\newpage 


\end{document}